\documentstyle[aps,preprint,tighten,floats,psfig]{revtex}
\begin{document}
\draft
\title{Probability of reflection by a random laser}
\author{C. W. J. Beenakker,$^{\rm a}$ J. C. J. Paasschens,$^{\rm a,b}$ and 
P. W. Brouwer$^{\rm a}$}
\address{$^{\rm a}$Instituut-Lorentz, University of Leiden,
P.O. Box 9506, 2300 RA Leiden, The Netherlands}
\address{$^{\rm b}$Philips Research Laboratories, 5656 AA Eindhoven, The
Netherlands}
\date{to appear in Phys.\ Rev.\ Lett.}
\maketitle
\begin{abstract}
A theory is presented (and supported by numerical simulations) for
phase-coherent reflection of light by a disordered medium which either absorbs
or amplifies radiation. The distribution of reflection eigenvalues is shown to
be the Laguerre ensemble of random-matrix theory. The statistical fluctuations
of the albedo (the ratio of reflected and incident power) are computed for
arbitrary ratio of sample thickness, mean free path, and absorption or
amplification length. On approaching the laser threshold all moments of the
distribution of the albedo diverge. Its modal value remains finite, however,
and acquires an anomalous dependence on the illuminated surface area.
\end{abstract}
\narrowtext

Recent experiments on turbid laser dyes \cite{Law94,Sha94,Wie95,Gen94} have
drawn attention to the remarkable properties of disordered media which are
optically active. The basic issue is to understand the interplay of
phase-coherent multiple scattering and amplification (or absorption) of
radiation. A quantity which measures this interplay is the albedo $a$, which is
the power reflected by the medium divided by the incident power. A thick
disordered slab which is optically passive has $a=1$. Absorption leads to $a<1$
and amplification to $a>1$. As the amplification increases the laser threshold
is reached, at which the average albedo becomes infinitely large \cite{Let67}.
Such a generator was referred to by its inventor V. S. Letokhov as a ``laser
with incoherent feedback'' \cite{Lav74}, because the feedback of radiation is
provided by random scattering and not by mirrors --- as in a conventional
laser.

The current renewed interest in random lasers owes much to the appreciation
that randomness is not the same as incoherence. Early theoretical work on this
problem was based on the equation of radiative transfer \cite{Ish78}, which
ignores phase coherence. Zyuzin \cite{Zyu94} and Feng and Zhang \cite{Fen95}
considered interference effects on the average albedo $\bar{a}$, averaged over
different configurations of the scattering centra. Their prediction of a
sharpening of the backscattering peak in the angular distribution of the
average reflected intensity has now been observed \cite{Wie95}. The other basic
interference effect is the appearance of large, sample-specific fluctuations of
the albedo around its average. These diverge faster than the average on
approaching the laser threshold \cite{Zyu95}, so that $\bar{a}$ is no longer
characteristic for the albedo of a given sample. In the present paper we will
show that, while all moments of the distribution function $P(a)$ of the albedo
diverge at the laser threshold, its modal value $a_{\rm max}$ remains finite.
The modal value is the value of $a$ at which $P(a)$ is maximal, and hence it is
the most probable value measured in a single experiment. The diagrammatic
perturbation theory of Refs.\ \cite{Zyu94,Fen95,Zyu95} can only give the first
few moments of $a$, and hence can not determine $a_{\rm max}$. Here we develop
a non-perturbative random-matrix theory for the entire distribution of the
reflection matrix, from which $P(a)$ can be computed directly.

We contrast the two cases of absorption and amplification. In the case of
absorption, $P(a)$ is a Gaussian with a width $\delta a$ smaller than the
average $\bar{a}$ by a factor $\sqrt{N}$, where $N\simeq S/\lambda^{2}\gg 1$ is
the number of modes associated with an illuminated area $S$ and wavelength
$\lambda$. In the case of amplification, both $\delta a$ and $\bar{a}$ increase
strongly on approaching the laser threshold --- in a manner which we compute
precisely. Below threshold, the mean and modal value of $a$ coincide. Above
threshold, the mean is infinite while the modal value is found to be
\begin{equation}
a_{\rm max}=1+0.8\,\gamma N.\label{amaxresult}
\end{equation}
Here $\gamma$ denotes the amplification per mean free path, assumed to be in
the range $N^{-2}\ll\gamma\ll 1$. The existence of a finite $a_{\rm max}$ is
due to the finiteness of the number of modes $N$ in a surface area $S$ (ignored
in radiative transfer theory). Since $a_{\rm max}$ scales with $N$ and hence
with $S$, and the incident power scales with $S$, it follows that the reflected
power scales {\em quadratically\/} rather than {\em linearly\/} with the
illuminated area. We suggest the name ``superreflection'' for this phenomenon.
To measure the albedo in the unstable regime above the laser threshold we
propose a time-resolved experiment, consisting of illumination by a short
intense pulse to pump the medium beyond threshold, rapidly followed by a weak
pulse to measure the reflected intensity before spontaneous emission has caused
substantial relaxation.

Our work on this problem was motivated by a recent paper by Pradhan and Kumar
\cite{Pra94} on the case $N=1$ of a single-mode waveguide. We discovered the
anomalous scaling with area in an attempt to incorporate the effects of
mode-coupling into their approach.

\begin{figure}
\hspace*{\fill}
\psfig{figure=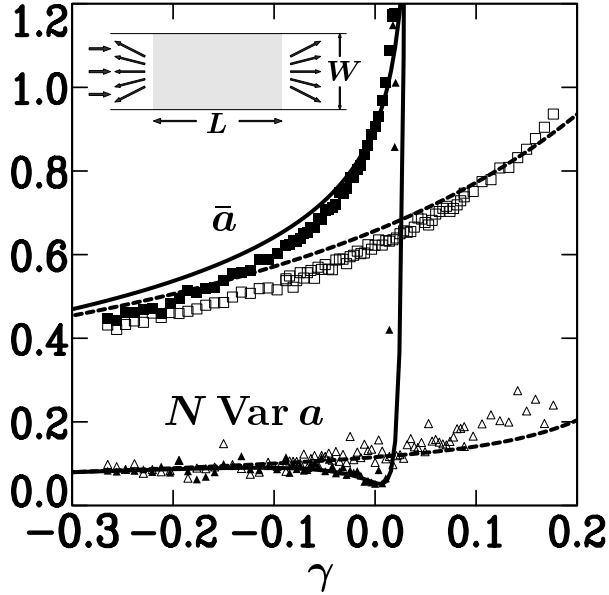,width=
8cm,bbllx=109pt,bblly=93pt,bburx=465pt,bbury=451pt}
\hspace*{\fill}
\caption[]{
Comparison between theory and simulation of the average albedo $\bar{a}$ (upper
curves, squares) and the variance ${\rm Var}\,a$ (lower curves, triangles) for
$L/l=1.92$ (dashed curves, open markers) and $L/l=9.58$ (solid curves, filled
markers). Negative $\gamma$ corresponds to absorption, positive $\gamma$ to
amplification. The curves are the theoretical result
(\protect\ref{baraVarasol}). The data points are a numerical simulation of a
two-dimensional lattice ($L=50d$ and $250d$, $W=51d$, $N=21$), averaged over
100 realizations of the disorder. The inset shows schematically the system
under consideration.
\label{baraVara_fig}
}
\end{figure}

We consider the reflection of a monochromatic plane wave (frequency $\omega$,
wavelength $\lambda$) by a slab (thickness $L$, area $S$) consisting of a
disordered medium (mean free path $l$) which either amplifies or absorbs the
radiation. We denote by $\sigma$ the amplification per unit length, a negative
value of $\sigma$ indicating absorption. The parameter $\gamma=\sigma l$ is the
amplification (or absorption) per mean free path. We treat the case of a scalar
wave amplitude, and leave polarization effects for future study. A discrete
number $N$ of scattering channels is defined by imbedding the slab in an
optically passive waveguide without disorder (see Fig.\ \ref{baraVara_fig},
inset). The number $N$ is the number of modes which can propagate in the
waveguide at frequency $\omega$. The $N\times N$ reflection matrix $r$ contains
the amplitudes $r_{mn}$ of waves reflected into mode $m$ from an incident mode
$n$. (The basis states of $r$ are normalized such that each carries unit
power.) The reflection eigenvalues $R_{n}$ ($n=1,2,\ldots N$) are the
eigenvalues of the matrix product $rr^{\dagger}$. The matrix $r$ is determined
by the $R_{n}$'s and by a unitary matrix $U$,
\begin{equation}
r_{mn}={\textstyle\sum_{k}}U_{mk}U_{nk}\sqrt{R_{k}}.\label{rURdef}
\end{equation}
Note that $r_{mn}=r_{nm}$ because of time-reversal symmetry. From $r$ one can
compute the albedo $a$ of the slab, which is the ratio of the reflected and
incident power:
\begin{equation}
a={\textstyle\sum_{m}}|r_{mn}|^{2}={\textstyle\sum_{k}}U_{nk}^{\vphantom{\ast}}
U_{nk}^{\ast}R_{k}.\label{adef}
\end{equation}

For a statistical description we consider an ensemble of slabs with different
configurations of scatterers. As in earlier work on optically passive media
\cite{Mel88a}, we make the isotropy assumption that $U$ is uniformly
distributed in the unitary group. This assumption breaks down if the transverse
dimension $W$ of the slab is much greater than its thickness $L$, but is
expected to be reasonable if $W\lesssim L$. As a consequence of isotropy, $a$
becomes statistically independent of the index $n$ of the incident mode. We
further assume that the wavelength $\lambda$ is much smaller than both the mean
free path $l$ and the amplification length $1/\sigma$. The evolution of the
reflection eigenvalues with increasing $L$ can then be described by a Brownian
motion process. To describe this evolution it is convenient to use the
parameterization
\begin{equation}
R_{n}=1+\mu_{n}^{-1},\;\;\mu_{n}\in(-\infty,-1)\cup(0,\infty).\label{mudef}
\end{equation}
The $L$-dependence of the distribution $P(\mu_{1},\mu_{2},\ldots\mu_{N})$ of
the $\mu$'s is governed by the Fokker-Planck equation
\begin{equation}
l\frac{\partial P}{\partial L}=\frac{2}{N+1}\sum_{i=1}^{N}
\frac{\partial}{\partial\mu_{i}}\mu_{i}(1+\mu_{i})\Bigl[\frac{\partial
P}{\partial\mu_{i}}+P\sum_{j\neq i}\frac{1}{\mu_{j}-\mu_{i}}
\mbox{}+\gamma(N+1)P\Bigr],\label{FP}
\end{equation}
with initial condition $\lim_{L\rightarrow 0}P=N\prod_{i}\delta(\mu_{i}+1)$. In
the single-channel case ($N=1$), the term $\sum_{j\neq i}$ is absent and Eq.\
(\ref{FP}) reduces to the differential equation studied by Pradhan and Kumar
\cite{Pra94,Ger59}. The multi-channel case is essentially different due to the
coupling of the eigenvalues by the term $\sum_{j\neq i}(\mu_{j}-\mu_{i})^{-1}$.
This term induces a {\em repulsion\/} of closely separated eigenvalues.
Equation (\ref{FP}) with $\gamma=0$ is known as the
Dorokhov-Mello-Pereyra-Kumar (DMPK) equation \cite{Dor82,Mel88b}, and has been
studied extensively in the context of electronic conduction \cite{Bee94b}. We
have generalized it to $\gamma\neq 0$, by adapting the approach of Ref.\
\cite{Mel88b} to a non-unitary scattering matrix.

The average $\bar{a}\equiv\langle a\rangle$ and the variance ${\rm
Var}\,a\equiv\langle(a-\bar{a})^{2}\rangle$ of the albedo (\ref{adef}) can be
computed by first averaging $U$ over the unitary group and then evaluating
moments of the $R_{k}$'s by means of Eq.\ (\ref{FP}) \cite{Mel91}. In the limit
$N\rightarrow\infty$ we obtain the differential equations
\begin{mathletters}
\label{baraVara}
\begin{eqnarray}
&&l\frac{d}{dL}\bar{a}=(\bar{a}-1)^{2}+2\gamma\bar{a},\label{bara}\\
&&l\frac{d}{dL}{\rm Var}\,a=4(\bar{a}-1+\gamma){\rm
Var}\,a+2N^{-1}\,\bar{a}(\bar{a}-1)^{2}.\label{Vara}
\end{eqnarray}
\end{mathletters}%
Corrections are smaller by a factor $|\gamma N^{2}|^{-1/2}$, which we assume to
be $\ll 1$. Eq.\ (\ref{bara}) for the average albedo is an old result of
radiative transfer theory \cite{Sel74}. Eq.\ (\ref{Vara}) for the variance is
new. It describes the sample-specific fluctuations of the albedo due to
interference of multiply scattered waves. Integration of Eq.\ (\ref{baraVara})
yields
\widetext
\begin{mathletters}
\label{baraVarasol}
\begin{eqnarray}
\bar{a}&=&1-\gamma+(2\gamma-\gamma^{2})^{1/2}\tan t,\label{barasol}\\
{\rm Var}\,a&=&(8N\cos^{4}t)^{-1}\Bigl(4\gamma(1-2\gamma)L/l
+2\gamma(1+\gamma)-4\gamma^{2}\cos 2t+2\gamma(1-\gamma)\cos 4t\nonumber\\
&&\mbox{}+(2-\gamma)^{-1}(2\gamma-\gamma^{2})^{1/2}\bigl[4\gamma(1-\gamma)\sin
2t-(1-4\gamma+2\gamma^{2})\sin 4t\bigr]\Bigr).\label{Varasol}
\end{eqnarray}
\end{mathletters}%
\narrowtext
\noindent
We have abbreviated $t=(2\gamma-\gamma^{2})^{1/2}\,L/l-\arcsin(1-\gamma)$.

Plots of Eq.\ (\ref{baraVarasol}) as a function of $\gamma$ are shown in Fig.\
\ref{baraVara_fig}, for two values of $L/l$. (The data points are numerical
simulations, discussed later.) In the case of absorption ($\gamma<0$), the
large-$L$ limit
\begin{mathletters}
\label{baraVarainfty}
\begin{eqnarray}
&&\bar{a}_{\infty}=1-\gamma-(\gamma^{2}-2\gamma)^{1/2},\label{barainfty}\\
&&{\rm Var}\,a_{\infty}=\frac{1}{2N}\,
\frac{\bar{a}_{\infty}(1-\bar{a}_{\infty})^{2}}
{1-\gamma-\bar{a}_{\infty}},\label{Varainfty}
\end{eqnarray}
\end{mathletters}%
can be obtained directly from Eq.\ (\ref{baraVara}) by equating the
right-hand-side to zero. The limit (\ref{baraVarainfty}) is reached when
$L/l\gg(\gamma^{2}-2\gamma)^{-1/2}$. In the case of amplification ($\gamma>0$),
Eq.\ (\ref{baraVarasol}) holds for $L$ smaller than the critical length
\begin{equation}
L_{\rm c}=l(2\gamma-\gamma^{2})^{-1/2}\arccos(\gamma-1)\label{Lcdef}
\end{equation}
at which $\bar a$ and ${\rm Var}\,a$ diverge. This is the laser threshold
\cite{Let67,Sel74}. For $\gamma<0$ the large-$L$ limit of the probability
distribution $P(a)$ of the albedo is well described by a Gaussian, with mean
and variance given by Eq.\ (\ref{baraVarainfty}). (The tails are non-Gaussian,
but carry negligible weight.) The modal value $a_{\rm max}$ of the albedo
equals $\bar{a}$. For $\gamma>0$ the large-$L$ limit of $P(a)$ can not be
reconstructed from its moments, but needs to be determined directly. We will
see that while $\bar{a}$ diverges, $a_{\rm max}$ remains finite.

\begin{figure}
\hspace*{\fill}
\psfig{figure=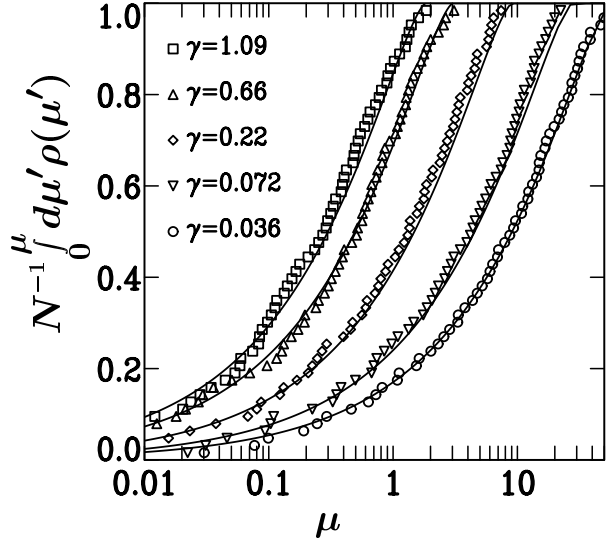,width= 8cm,bbllx=58pt,bblly=95pt,bburx=449pt,bbury=449pt}
\hspace*{\fill}
\caption[]{
Comparison between theory and simulation of the cumulative density of the
variables $\mu_{n}$ (related to the reflection eigenvalues by
$R_{n}=1+\mu_{n}^{-1}$). Curves are computed from the density
(\protect\ref{rhomu}) of the Laguerre ensemble; Data points are from the
simulation ($L=500d=19.2\,l$, $W=151d$, $N=63$), for a single realization of
the disorder.
\label{rhomu_fig}
}
\end{figure}

The large-$L$ limit $P_{\infty}(\mu_{1},\mu_{2},\ldots\mu_{N})$ of the
distribution of the $\mu$'s is obtained by equating to zero the expression
between square brackets in Eq.\ (\ref{FP}). The result is
\begin{equation}
P_{\infty}=C\prod_{i}\exp\Bigl(-\gamma(N+1)\mu_{i}\Bigr)
\prod_{i<j}|\mu_{j}-\mu_{i}|,\label{Pinfty}
\end{equation}
with $C$ a normalization constant. Eq.\ (\ref{Pinfty}) holds for both positive
and negative $\gamma$, but the support of $P_{\infty}$ depends on the sign of
$\gamma$: All $\mu$'s have to be $>0$ for $\gamma>0$ (amplification) and $<-1$
for $\gamma<0$ (absorption). In what follows we take $\gamma>0$. The function
(\ref{Pinfty}) is known in random-matrix theory as the distribution of the {\em
Laguerre ensemble} \cite{Bro65}. The density
$\rho(\mu)=\langle\sum_{i}\delta(\mu-\mu_{i})\rangle$ of the $\mu$'s is a
series of Laguerre polynomials, hence the name. For $\gamma N^{2}\gg 1$ one has
asymptotically
\begin{equation}
\rho(\mu)=(N/\pi)\left(2\gamma/\mu-\gamma^{2}\right)^{1/2},
\;\;0<\mu<2/\gamma.\label{rhomu}
\end{equation}
The square-root singularity at $\mu=0$ is cut off in the exact density
\cite{Nag93}, such that $\rho=\gamma N^{2}$ if $\mu\lesssim 1/\gamma N^{2}$.
The cumulative density is plotted in Fig.\ \ref{rhomu_fig}, together with the
numerical simulations (discussed below).

\begin{figure}
\hspace*{\fill}
\psfig{figure=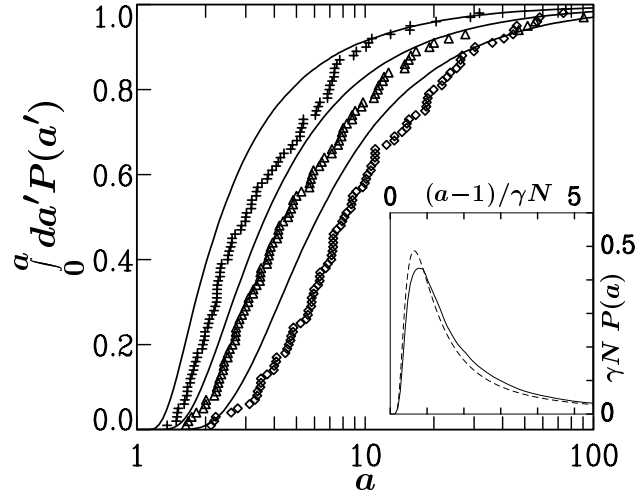,width= 8cm,bbllx=58pt,bblly=95pt,bburx=478pt,bbury=450pt}
\hspace*{\fill}
\caption[]{
Comparison between theory and simulation of the cumulative probability
distribution of the albedo ($L=500d=19.2\,l$, $\gamma=0.07$). Solid curves are
obtained by numerically averaging over the Laguerre ensemble; Data points are
the results of the simulation, averaged over 100 realizations of the disorder.
The three sets of data are for $W=25d$, $N=10$ (plusses), $W=51d$, $N=21$
(triangles), and $W=101d$, $N=42$ (diamonds). The inset compares the continuum
approximation (\protect\ref{Paresult}) for $P(a)$ (dashed) with the exact
large-$N$ limit of the Laguerre ensemble (solid).
\label{Paresult_fig}
}
\end{figure}

We seek the probability distribution of the albedo
\begin{equation}
P(a)=\left\langle\delta\bigl(a-1-{\textstyle\sum_{k}}
U_{nk}^{\vphantom{\ast}}U_{nk}^{\ast}\mu_{k}^{-1}\bigr)\right\rangle.
\label{Padef}
\end{equation}
The average $\langle\cdots\rangle$ consists of the average of $U$ over the
unitary group followed by the average of the $\mu_{k}$'s over the Laguerre
ensemble. The averages can be done analytically for $N^{-2}\ll\gamma\ll 1$ (in
the continuum approximation \cite{Dys62}, i.e.\ by ignoring the discreteness of
the eigenvalues), and numerically for any $N,\gamma$ (by Monte Carlo
integration, i.e.\ by randomly sampling the Laguerre ensemble). The analytical
result is an inverse Laplace transform,
\begin{mathletters}
\label{Paresult}
\begin{equation}
P(a)=\frac{1}{2\gamma N}\int_{-{\rm i}\infty}^{{\rm i}\infty}
\frac{ds}{2\pi{\rm i}}\,\exp\bigl[\case{1}{2}s(a-1)/\gamma
N-2f(s)\bigr][1+\case{1}{4}f(s)]^{2},\label{Paresulta}
\end{equation}
where $f$ is an implicit function of the Laplace variable $s$:
\begin{equation}
(s-\case{1}{2}f+\case{1}{2}\sqrt{4f+f^2})^{-1/2}+ 2(f-\sqrt{4f+f^2})^{-1}
+1=0.\label{fsrelation}
\end{equation}
\end{mathletters}%
The continuum approximation (\ref{Paresult}) is plotted in the inset of Fig.\
\ref{Paresult_fig} (dashed curve). It is close to the exact numerical large-$N$
result (solid curve). The modal value $a_{\rm max}$ of the albedo is given by
Eq.\ (\ref{amaxresult}). The distribution $P(a)$ drops off
$\propto\exp[-2\gamma N/(a-1)]$ for smaller $a$ and $\propto a^{-5/3}$ for
larger $a$, so that all moments diverge.

To test these predictions of random-matrix theory on a model system, we have
carried out numerical simulations of the analogous electronic Anderson model
with a complex scattering potential, using the recursive Green's function
technique \cite{Bar91}. The disordered medium is modeled by a two-dimensional
square lattice (lattice constant $d$, length $L$, width $W$). The relative
dielectric constant $\varepsilon=\varepsilon_{1}+{\rm i}\varepsilon_{2}$
(relative to the value outside the disordered region) has a real part
$\varepsilon_{1}$ which fluctuates from site to site between
$1\pm\delta\varepsilon$, and a non-fluctuating imaginary part
$\varepsilon_{2}$. The multiple scattering of a scalar wave $\Psi$ (wave number
$k=2\pi/\lambda$) is described by discretizing the Helmholtz equation
$(\nabla^{2}+k^{2}\varepsilon)\Psi=0$. The mean free path $l$ which enters in
Eq.\ (\ref{FP}) is obtained from the average albedo $\bar{a}=(1+l/L)^{-1}$
without amplification ($\varepsilon_{2}=0$). We choose $k^{2}=1.5\,d^{-2}$,
$\delta\varepsilon=1$, leading to $l=26.1\,d$. The parameter $\sigma$ (and
hence $\gamma=\sigma l$) is obtained from the analytical solution of the
discretized Helmholtz equation in the absence of disorder
($\delta\varepsilon=0$). The complex longitudinal wavenumber $k_{n}$ of
transverse mode $n$ then satisfies the dispersion relation
\begin{equation}
\cos(k_{n}d)+\cos(n\pi d/W)=2-\case{1}{2}(kd)^{2}(1+{\rm i}\varepsilon_{2}),
\label{dispersion}
\end{equation}
and leads to $\sigma=-2N^{-1}\,{\rm Im}\,\sum_{n}k_{n}$. The albedo
(\ref{adef}) is computed for normal incidence. Data points in Figs.\ 1--3 are
the numerical results. The agreement with the curves from random-matrix theory
is quite remarkable, given that there are {\em no adjustable parameters}.

In summary, we have presented a random-matrix theory for the reflection matrix
of a disordered medium with absorption or amplification. In the limit of a
semi-infinite medium, the distribution of the reflection eigenvalues is that of
the Laguerre ensemble. The corresponding distribution of the albedo is a
Gaussian in the case of absorption. In the case of amplification, the
distribution has diverging moments but a finite modal value. By ignoring
spontaneous emission of radiation, we could examine the light reflected in
response to an incident wave, separately from spontaneously generated light. In
future work, we intend to include source terms in our description, to account
for amplified spontaneous emission and the resulting relaxation of the unstable
state above the laser threshold. We also plan to apply our approach to
frequency-dependent fluctuations and to the case of diffusive, rather than
plane-wave illumination.

We thank A. Lagendijk, M. B. van der Mark, and D. S. Wiersma for helpful
discussions. This work was supported by the Dutch Science Foundation NWO/FOM.

\end{document}